\documentclass[aps,prb,preprint,amsmath,amssymb,showpacs,showkeys,superscriptaddress
]{revtex4}
\usepackage{graphicx}
\usepackage{dcolumn}
\usepackage{hyperref}
\newcommand{\be}{\begin{equation}}
\newcommand{\ee}{\end{equation}}
\newcommand{\ba}{\begin{eqnarray}}
\newcommand{\ea}{\end{eqnarray}}
\begin{document}
\title{Evaluation of phenomenological one-phase criteria \\
for the melting and freezing of softly repulsive particles}
\author{Franz Saija}
\altaffiliation{Corresponding author}
\email{\tt saija@me.cnr.it}
\affiliation{Istituto per i Processi Chimico-Fisici del CNR,
Sezione di Messina,\\Via La Farina 237, 98123 Messina, Italy}
\author{Santi Prestipino}
\email{\tt Santi.Prestipino@unime.it}
\affiliation{Universit\`a degli Studi di Messina,\\Dipartimento di
Fisica, Contrada Papardo, 98166 Messina, Italy}
\author{Paolo V. Giaquinta}
\email{\tt Paolo.Giaquinta@unime.it}
\affiliation{Universit\`a degli Studi di Messina,\\Dipartimento di
Fisica, Contrada Papardo, 98166 Messina, Italy}
\date{\today}

\begin{abstract} We test the validity of some widely used phenomenological criteria for the localization of the
fluid-solid transition thresholds against the phase diagrams of
particles interacting through the exp-6, inverse-power-law, and
Gaussian potentials. We find that one-phase rules give, on the
whole, reliable estimates of freezing/melting points. The agreement
is ordinarily better for a face-centered-cubic solid than for a
body-centered-cubic crystal, even more so in the presence of a
pressure-driven re-entrant transition of the solid into a denser
fluid phase, as found in the Gaussian-core model.
\end{abstract}

\maketitle

\section{Introduction} Predicting the phase diagram of a real substance is an issue of outstanding importance in
materials science, with manifest connections with both basic and
applied research. Over the last decades, many theoretical and
computational efforts have been devoted to deriving the phase
diagram of a variety of model systems, also with the aim at gaining
a more global perspective on the fundamental link between the
macroscopic phase behaviour and the underlying atomic or molecular
interactions~\cite{Young,Monson}. Fluid-solid phase transitions are
among the most studied discontinuous phase changes. With the advent
of advanced numerical-simulation methods for the calculation of free
energies, the accuracy of the estimated fluid-solid coexistence
boundaries has significantly improved~\cite{Monson}. Thermodynamic
equilibria can be easily computed once one knows the Gibbs free
energy of the competing phases. In fact, the stable macroscopic
phase of a substance at equilibrium is the one that minimizes the
appropriate thermodynamic potential for a given choice of extensive
and/or intensive parameters. However, calculating the free energy of
either a dense fluid or a hot solid still remains a demanding
computational task that requires intensive simulations to be carried
out at several state points as well as some preliminary selection of
the most likely candidate solid structures. For such reasons, a
number of empirical rules have been proposed since the early times
of statistical mechanics in an attempt to correlate phase-transition
thresholds with the thermodynamic or structural properties of the
solid and fluid phases, respectively~\cite{Monson,Lowen}. All such
criteria are typically based on the properties of one phase only
and, in general, can be easily implemented with a modest
computational effort.

The recent availability of more accurate results on the phase
diagrams of some reference models for effective pair interactions in
simple and complex fluids~\cite{Prestipino1,Prestipino2,Saija1}
provides an opportunity to test the reliability of some well known
one-phase criteria for the fluid-solid transition. We refer, in
particular, to the celebrated Lindemann~\cite{Lindemann} and
Hansen-Verlet~\cite{Hansen1} rules for melting and freezing,
respectively. In this paper, we shall also analyze the relative
performances of other two criteria, {\it viz.}, the
Ravech\'e-Mountain-Streett freezing rule~\cite{Raveche} and an
entropy-based ordering criterion originally proposed by Giaquinta
and Giunta for hard spheres~\cite{Giaquinta1} and later extended to
other continuum as well as lattice-gas models~\cite{Giaquinta2}.

\section{One-phase rules for the fluid-solid transition}
\subsection{Lindemann's melting law} The Lindemann ratio $L$ is defined as the root mean square displacement
of particles in a crystalline solid about their equilibrium lattice positions, divided by their nearest-neighbor distance $a$: \be
L=\frac{1}{a}\left\langle\frac{1}{N}\sum_{i=1}^N\left(\Delta{\bf R}_i\right)^2\right\rangle^{1/2}\,, \label{lindratio} \ee \noindent
where  $N$ is the number of particles and the brackets $\left\langle\cdots\right\rangle$ denote the average over the
dynamic trajectories of the particles. The Lindemann criterion states
that the crystal melts when $L$ overcomes some ``critical'' (yet not
specified {\em a priori}) value $L_c$~\cite{Lindemann}. Obviously,
one would hope this latter quantity to be approximately the same for different pair
potentials and thermodynamic conditions. In fact, the Lindemann
ratio is not universal at all, its values spanning in the range $0.12 -
0.19$. More specifically, $L_c$ is reported to be $0.15 - 0.16$ in a
face-centered-cubic (FCC) solid  and $0.18 - 0.19$ in a body-centered-cubic (BCC) solid (see, {\it e.g.},~\cite{Meijer}).

\subsection{Hansen-Verlet freezing rule} The Hansen-Verlet rule is a statement on the height, $S_{\rm max}$, of the
first peak of the liquid structure factor at freezing. According to
this criterion, $S_{\rm max} \simeq 2.85$ along the freezing curve
of simple fluids ~\cite{Hansen1}. Indeed, also the value of $S_{\rm
max}$ at freezing has been found to depend on the softness of the
potential as well as on the dimensionality of the hosting
space~\cite{Hansen2}.

\subsection{Ravech\'e-Mountain-Streett freezing criterion}
Ravech\'e, Mountain, and Streett proposed an empirical criterion for the
freezing of a classical Lennard-Jones system that is based, instead, on the
radial distribution function (RDF) of the liquid~\cite{Raveche}. They
focused on the ratio between the values of the RDF at distances
corresponding to the first non-zero minimum and to the largest
maximum, $\Gamma=g(r_{\rm min})/g(r_{\rm max})$. As the Authors actually noted, one should expect that the magnitude of the maxima and minima of
the RDF are not entirely arbitrary as the area under $g(r)$ is proportional to the number of nearest neighbors which is
fixed for a given lattice. Ravech\'e, Mountain, and Streett indicated
$0.20\pm 0.02$ as the value of $\Gamma$ at freezing.

\subsection{An entropy-based ordering criterion} The residual multiparticle entropy (RMPE) of
a homogeneous and isotropic fluid \be \Delta s=s_{\rm ex}-s_2\,
\label{ds} \ee is the difference between the excess entropy per
particle, $s_{\rm ex}$, and the integrated contribution of pair
density correlations to the entropy of the fluid~\cite{Nettleton}:
\be s_2=-\frac{1}{2}\rho\int{\rm d}{\bf r}\,\left[g(r)\ln
g(r)-g(r)+1\right]\,, \label{s2} \ee where $\rho$ is the number
density. In Eqs.~2 and 3, entropies are given in units of the
Boltzmann constant, $k_{\rm B}$. The RMPE has been found to
vanish~\cite{Giaquinta1,Giaquinta2} whenever a disordered (or, even,
partially ordered) fluid transforms into a more structured phase, in
both two and three dimensions~\cite{Saija2,Saija3}. In fact, this
criterion is not restricted to the fluid-solid phase transition but
also yields reliable information on the location of other
first-order phase transitions, such as the phase separation of
binary mixtures~\cite{Saija4} and the formation of mesophases
(nematic, smectic) in liquid crystals~\cite{Costa}. Moreover, at
variance with other phenomenological rules, the zero-RMPE criterion
is a self-contained statement whose implementation does not hinge
upon an external, context-dependent input. The relation of this
criterion with the Hansen-Verlet rule was discussed
in~\cite{Saija3}.

\section{Interaction models}
\subsection{The modified Buckingam potential}
It is usually thought that the thermal behaviour of rare gases is
well accounted for by the Lennard-Jones potential. However, it turns
out that very dense rare gases are better described by a pair
potential with a softer repulsive shoulder. In this respect, a
better choice is the modified Buckingam potential, also known as the
exp-6 potential: \be v_{\rm B}(r)=\left\{
\begin{array}{ll}
+\infty\,, & r<\sigma \\
\frac{\epsilon}{\alpha-6}
\left\{6\exp\left[\alpha\left(1-\frac{r}{r_{\rm min}}\right)\right]-
\alpha\left(\frac{r_{\rm min}}{r}\right)^6\right\}\,, & r\geq\sigma
\end{array}
\right. \,, \label{exp6} \ee where $\alpha>6$ controls the softness
of the repulsion and $\epsilon>0$ is the depth of the potential
minimum at $r_{\rm min}$. The value of $\sigma$ in Eq.\,(\ref{exp6})
is such that, for assigned values of $\alpha$ and $r_{\rm min}$, the
function in the second line of Eq.\,(\ref{exp6}) attains its maximum
at $r=\sigma$. Ross and McMahan showed that, for a suitable choice
of the parameters $\alpha$, $\epsilon$, and $r_{\rm min}$, $v_{\rm
B}(r)$ yields a faithful representation of many thermodynamic
properties of heavy rare gases at high densities~\cite{Ross1}.

At extreme thermodynamic conditions (very high pressure and
temperature), the structure of the fluid is largely determined by
the short-range repulsive part of the potential. A recent
computational study of the exp-6 potential, with parameters
appropriate for Xenon, has shown that, at low temperatures and for pressures as large as 60 GPa, the
stable crystalline phase is a FCC solid~\cite{Saija1}. However, for still larger pressures, a BCC
phase shows up, over a narrow range of temperatures, between the
fluid and the FCC phase. This finding is consistent with recent
diamond-anvil-cell data on Xenon~\cite{Ross2}.

The Hansen-Verlet criterion has been tested against the low-pressure
part of the exp-6 freezing line~\cite{Vortler,Lisal}. As anticipated
in Sec.\,2, $S_{\rm max}$ was found to depend at freezing on
$\alpha$, its value decreasing from $3.5$ to $2.85$ as $\alpha$
increases from $11.5$ to $14.5$.

\subsection{Inverse-power-law repulsive potentials}
Inverse-power-law (IPL) potentials \be v_{\rm
I}(r)=\epsilon\left(\frac{\sigma}{r}\right)^n\, \label{ipl} \ee
provide a continuous path from hard spheres ($n\rightarrow\infty$)
to the one-component plasma ($n=1$)~\cite{Hoover,Laird,Agrawal}.
This model has also been used to describe the effective interatomic
repulsion $(\epsilon>0)$ in a metal subject to extreme thermodynamic
conditions. Once $n$ is fixed, the thermodynamic properties of the
model can be expressed in terms of one single quantity
$\gamma=\rho^*T^{*\,-3/n}$, $\rho^*=\rho\sigma^3$ and $T^*=k_{\rm
B}T/\epsilon$ being the reduced density and temperature,
respectively. For large values of $n$, the BCC phase is unstable
with respect to shear modes and the fluid freezes into a FCC solid.
As $n$ decreases, the potential becomes increasingly softer and
longer ranged until, for $n\approx 8$, the BCC phase becomes
mechanically stable. For smaller values of $n$, the entropy of the
BCC solid turns out to be larger than that of the FCC phase in the
freezing region. Upon reducing $\gamma$, a FCC-BCC transition
becomes possible before melting. Exact free-energy calculations have
recently confirmed this scenario~\cite{Prestipino2}. According to
this study, the BCC phase is thermodynamically stable for values of
the inverse-power exponent less than $7.1$.

\subsection{The repulsive Gaussian potential} The Gaussian-core model (GCM) describes a system of particles interacting
through the bounded repulsive potential~\cite{Stillinger1}: \be v_{\rm G}(r)=\epsilon\exp(-r^2/\sigma^2)\,. \label{gcm}
\ee This potential has been used to represent the effective entropic repulsion originated by the self-avoidance of beads
in a dilute dispersion of polymers. Nothwistanding its apparent simplicity, the GCM has a rich phase
diagram~\cite{Lang,Prestipino1,Prestipino2}. A peculiarity of this model is that no solid phase is stable for
temperatures above $k_{\rm B}T_{\rm max}/\epsilon=0.00874$. A re-entrant melting behavior is observed below this
temperature. Upon compression and for temperatures in the range $0.0038 < k_{\rm B}T/\epsilon < 0.00874$, the
low-density fluid freezes into a BCC solid which eventually melts at higher densities. At lower temperatures, $0.0031 <
k_{\rm B}T/\epsilon < 0.0038$, the BCC phase is stable at low densities over a tiny region. In fact, upon compression,
it soon transforms into a FCC phase to re-enter the phase diagram at higher densities, before melting. For $k_{\rm
B}T/\epsilon < 0.0031$, the sequence of phases exploited by the GCM with increasing densities is just
fluid-FCC-BCC-fluid.

\section{Monte Carlo simulation}
We performed canonical Monte Carlo (MC) simulations of the three
models presented in Sec.~3, using the standard Metropolis algorithm
for the sampling of the equilibrium distribution function in
configurational space. The number of particles $N$ was chosen so as
to fit the chosen crystalline structure in a cubic simulation box
with an integer number of cells, specifically: $N=4m^3$ for a FCC
solid and $N=2m^3$ for a BCC solid, $m$ being the number of cells
along any spatial direction. Our samples consisted of $1372$
particles for the FCC solid (as well as for the fluid phase), and of
$1458$ particles for the BCC solid. Such sizes are large enough to
ensure that finite-size corrections of the quantities we are
interested in are typically smaller than numerical
errors~\cite{Prestipino2}.

Periodic conditions were applied to the cell boundaries. For given
$N$, the length $\ell$ of the box was chosen so as to fulfill the
density constraint, namely $\ell=(N/\rho)^{1/3}$. The distance $a$
between two nearest-neighbour lattice sites was
$(\sqrt{2}/2)(\ell/m)$ for a FCC crystal and $(\sqrt{3}/2)(\ell/m)$
for a BCC crystal. For each $\rho$ and $T$, the equilibration of the
sample typically took $2\times 10^3$ MC sweeps, a sweep consisting
of one attempt to sequentially change the positions of all the
particles. The relevant thermodynamic averages were computed over a
trajectory whose length ranged between $2\times 10^4$ and $6\times
10^4$ sweeps. The excess energy per particle $u_{\rm ex}$, the
pressure $P$, and (in the solid phase) the mean square deviation
$\left\langle (\Delta{\bf R})^2 \right\rangle$ of a particle from
its reference lattice position were especially monitored. We
computed the RDF of the fluid up to a distance $\ell/2$ and
calculated the structure factor \be S(q)=1+\rho\int{\rm d}{\bf
r}\,\exp(-{\rm i}{\bf q}\cdot{\bf r})\left[g(r)-1\right]\,.
\label{strufact} \ee More technical details about the estimate of
statistical errors affecting thermal averages and the computation of
free energies can be found in the original
references~\cite{Saija1,Prestipino1,Prestipino2}. We just note here
that the value of $s_{\rm ex}$ follows immediately once we know the
energy and the free energy of the system in a given thermodynamic
state.

In principle, when calculating the Lindemann ratio, one may
encounter a technical problem arising from particles that jump from
one lattice site to another. In this case, one has the problem of
deciding which lattice site the position of a given particle should
be referred to. In practice, such an exchange is very rare and was
never observed in our simulations but for the IPL potential model
with $n=5$. We finally note that, in order to take care of the small
drift of the system's center of mass after each accepted MC move,
the reference lattice was shifted by the same amount the center of
mass of the particles had diffused during the move.

\section{Results}
The values of the Lindemann and Ravech\'e-Mountain-Streett ratios,
calculated at a number of points along the melting/freezing lines of
the three interaction models, are given in Tables I -- III. The
estimated numerical accuracy of such values does not exceed a few
units in the third decimal figure. We found that the Lindemann ratio
is very close to 0.15 at the melting point of a FCC solid; it is
larger, approximately $0.18$, for a BCC structure. The only current
exception is the re-entrant melting of the GCM where $L_c\simeq
0.16$, notwithstanding the BCC nature of the melting solid. Our
results for the Ravech\'e-Mountain-Streett ratios are close to the
reference value ($0.20$) reported in the literature but, again, for
the fluid-solid transition undergone by the GCM model at high
densities that is characterized by larger values of $\Gamma$, in the
range $0.23 - 0.24$.

The freezing thresholds predicted by the Hansen-Verlet rule and by
the entropy-based ordering criterion are drawn in Figs.\,1 -- 3.
Overall, the agreement with the numerical simulation data is
satisfactory. The Hansen-Verlet rule almost systematically
underestimates the freezing density while the entropy-based
criterion overestimates the range of fluid stability. The distance
between the predictions given by the two criteria decreases with the
temperature in the exp-6 model and with the steepness of the
potential -- i.e., for increasing values of the power exponent --
when the particles interact through an IPL potential. As far as this
latter model is concerned, we recall that in the limit $1/n
\rightarrow 0$ one does actually recover hard spheres which freeze
at a reduced density $\rho \sigma^3 = 0.94$. At this density the
height of the first peak of the structure factor of hard spheres is
precisely $2.85$, the value on which Hansen and Verlet based, {\it a
posteriori}, their solidification rule for simple
fluids~\cite{Hansen1}. Correspondingly, Giaquinta and Giunta noted
that the RMPE of hard spheres vanishes at freezing, an observation
that gave rise to the entropy-based ordering
criterion~\cite{Giaquinta1}.

As for the GCM, the predictions of both criteria are extremely
accurate at low temperature and density, all along the fluid-FCC
border. However, the Hansen-Verlet rule works better along the
fluid-BCC border, in a region where the fluid phase eventually shows
non-conventional features associated with the re-entrant melting
phenomenon.

\section{Concluding remarks}
In this paper we tested the predictions of a few popular one-phase
criteria, frequently used to estimate melting and freezing points,
on some classical reference systems for the liquid state: the
modified Buckingam potential, the inverse-power-law potential, and
the Gaussian core model. Phenomenological rules, such as those
discussed in this paper, cannot replace the proper thermodynamic
prescriptions for the coexistence of liquid and solid phases at
equilibrium. However, also the present analysis confirms that their
predictions are usually sound and reliable. In some cases, the
agreement of such empirical estimates with the rigorous indications
provided by free-energy calculations for the coexisting phases is
more than qualitative. We conclude that approximate rules based on
thermodynamical, structural, or dynamical properties of one phase
only can be very helpful in gaining, with a low computational cost,
preliminary information on the location of the fluid-solid phase
boundaries of a given substance. Such a positive assessment on the
overall reliability of one-phase criteria, that is coherently
supported by many independent studies on a variety of model systems,
may also give confidence in the use of some of them -- specifically,
the freezing criteria, estimating the stability threshold of the
disordered phase --, even when the crystalline structure of the
coexisting solid cannot be easily anticipated.

\newpage
%
%
\begin{center}
\large
FIGURE CAPTIONS
\normalsize
\end{center}
\begin{description}
\item[{\bf Fig.\,1 :}]
Phase diagram of the exp-6 model potential: The lines mark
the stability limit for each homogeneous phase~\cite{Saija1}.
Open and solid circles indicate the
freezing thresholds predicted by the Hansen-Verlet rule and by the
entropy-based criterion, respectively.

\item[{\bf Fig.\,2 :}]
Phase diagram of the inverse-power-law model potential: The lines
mark the stability limit for each homogeneous
phase~\cite{Prestipino1}. Open and solid circles indicate the
freezing thresholds predicted by the Hansen-Verlet rule and by the
entropy-based criterion, respectively.

\item[{\bf Fig.\,3 :}]
Phase diagram of the Gaussian-core model: The lines mark the
stability limit for each homogeneous phase~\cite{Prestipino2}. Open
and solid circles indicate the freezing thresholds predicted by the
Hansen-Verlet rule and by the entropy-based criterion, respectively.
\end{description}
\newpage
%
%
\begin{table}
\caption{Lindemann and Ravech\'e-Mountain-Streett ratios in the
exp-6 model, calculated for a number of states along the
high-temperature part of the melting line. The solid is FCC for
$T^*=4.25,8.15,12.77$, and 16, BCC otherwise.}
\begin{tabular}{cccc} \\
\hline
$T^*$ & $\rho^*$ & $L_c$ & $\Gamma$ \\
\hline
$4.25$  & $2.055$ & $0.153$ & $0.195$ \\
$8.15$  & $2.571$ & $0.148$ & $0.194$ \\
$12.77$ & $3.022$ & $0.153$ & $0.194$ \\
$16$    & $3.295$ & $0.152$ & $0.192$ \\
$20$    & $3.601$ & $0.180$ & $0.195$ \\
$25$    & $3.958$ & $0.180$ & $0.197$ \\
\hline
\end{tabular}
\end{table}
%
%
\begin{table}
\caption{ Lindemann and Ravech\'e-Mountain-Streett ratios in the IPL
model, calculated for a number of values of the power exponent $n$.
The solid coexisting with the fluid is FCC for $1/n=0.10,0.12$, and
0.14, BCC otherwise.}
\begin{tabular}{cccc} \\
\hline
$1/n$ & $\rho^*$ & $L_c$ & $\Gamma$ \\
\hline
$0.10$ & $1.327$ & $0.151$ & $0.188$ \\
$0.12$ & $1.534$ & $0.154$ & $0.195$ \\
$0.14$ & $1.817$ & $0.155$ & $0.193$ \\
$0.15$ & $1.968$ & $0.165$ & $0.199$ \\
$0.16$ & $2.179$ & $0.180$ & $0.200$ \\
$0.17$ & $2.387$ & $0.181$ & $0.204$ \\
\hline
\end{tabular}
\end{table}
%
%
\begin{table}
\caption{Lindemann and Ravech\'e-Mountain-Streett ratios calculated
along the GCM freezing line. The solid coexisting with the fluid is
FCC for the first two entries only, BCC otherwise.}
\begin{tabular}{cccc} \\
\hline
$T^*$ & $\rho^*$ & $L_c$ & $\Gamma$ \\
\hline
$0.0020$ & $0.0800$ & $0.151$ & $0.185$ \\
$0.0030$ & $0.0938$ & $0.152$ & $0.202$ \\
$0.0033$ & $0.0978$ & $0.181$ & $0.185$ \\
$0.0035$ & $0.0998$ & $0.181$ & $0.196$ \\
$0.0037$ & $0.1031$ & $0.180$ & $0.202$ \\
$0.0038$ & $0.1038$ & $0.180$ & $0.207$ \\
$0.0040$ & $0.1074$ & $0.178$ & $0.187$ \\
$0.0060$ & $0.1332$ & $0.178$ & $0.210$ \\
$0.0080$ & $0.1792$ & $0.173$ & $0.208$ \\
\hline \hline
$0.0020$ & $0.6855$ & $0.162$ & $0.238$ \\
$0.0040$ & $0.5219$ & $0.163$ & $0.232$ \\
$0.0060$ & $0.4208$ & $0.165$ & $0.235$ \\
$0.0080$ & $0.3158$ & $0.165$ & $0.229$ \\
\hline
\end{tabular}
\end{table}

\newpage
%
%
\begin{figure}
\includegraphics[width=12cm,angle=0]{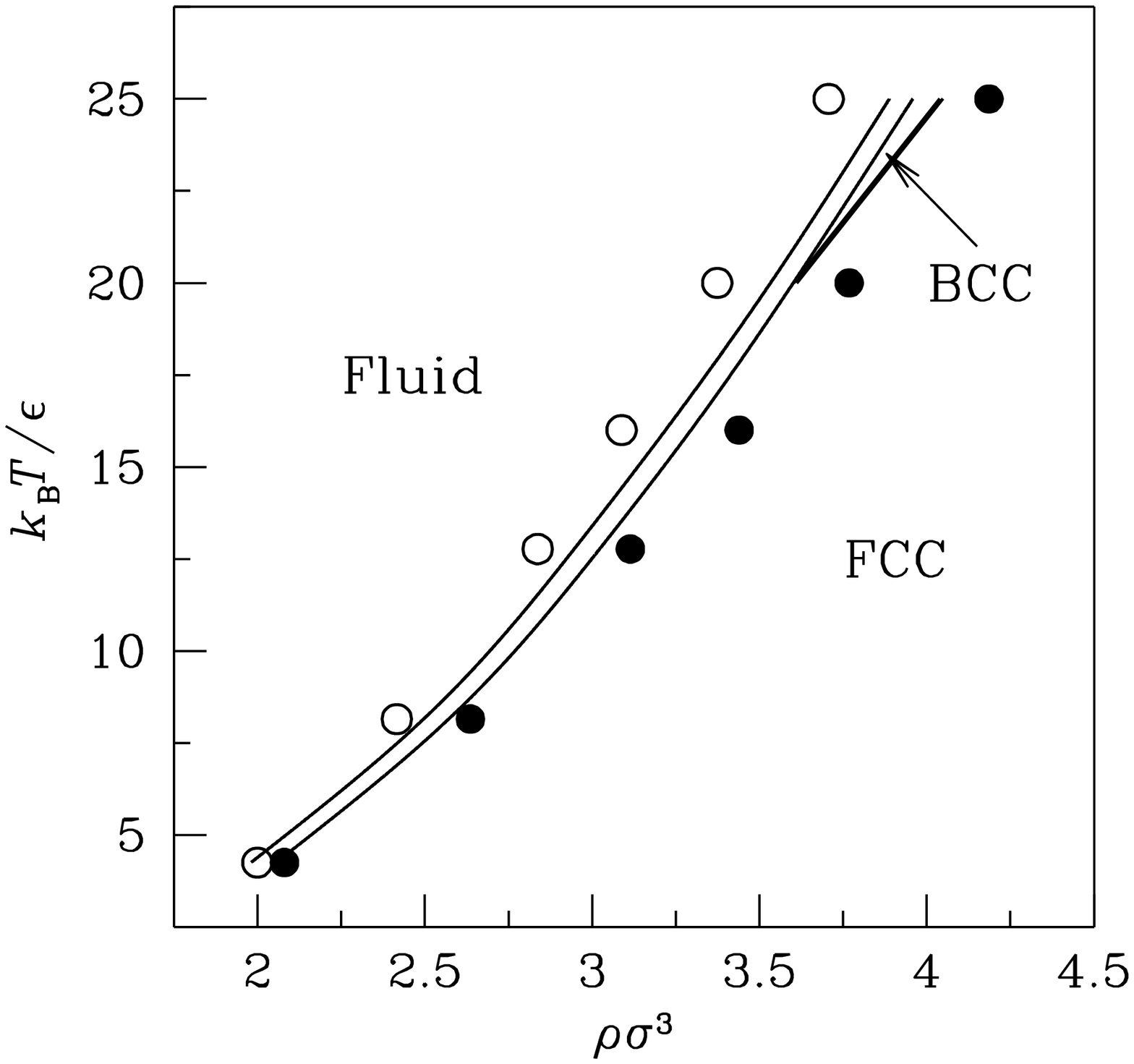}
\caption{} \label{fig1}
\end{figure}
%
%
\begin{figure}
\includegraphics[width=12cm,angle=0]{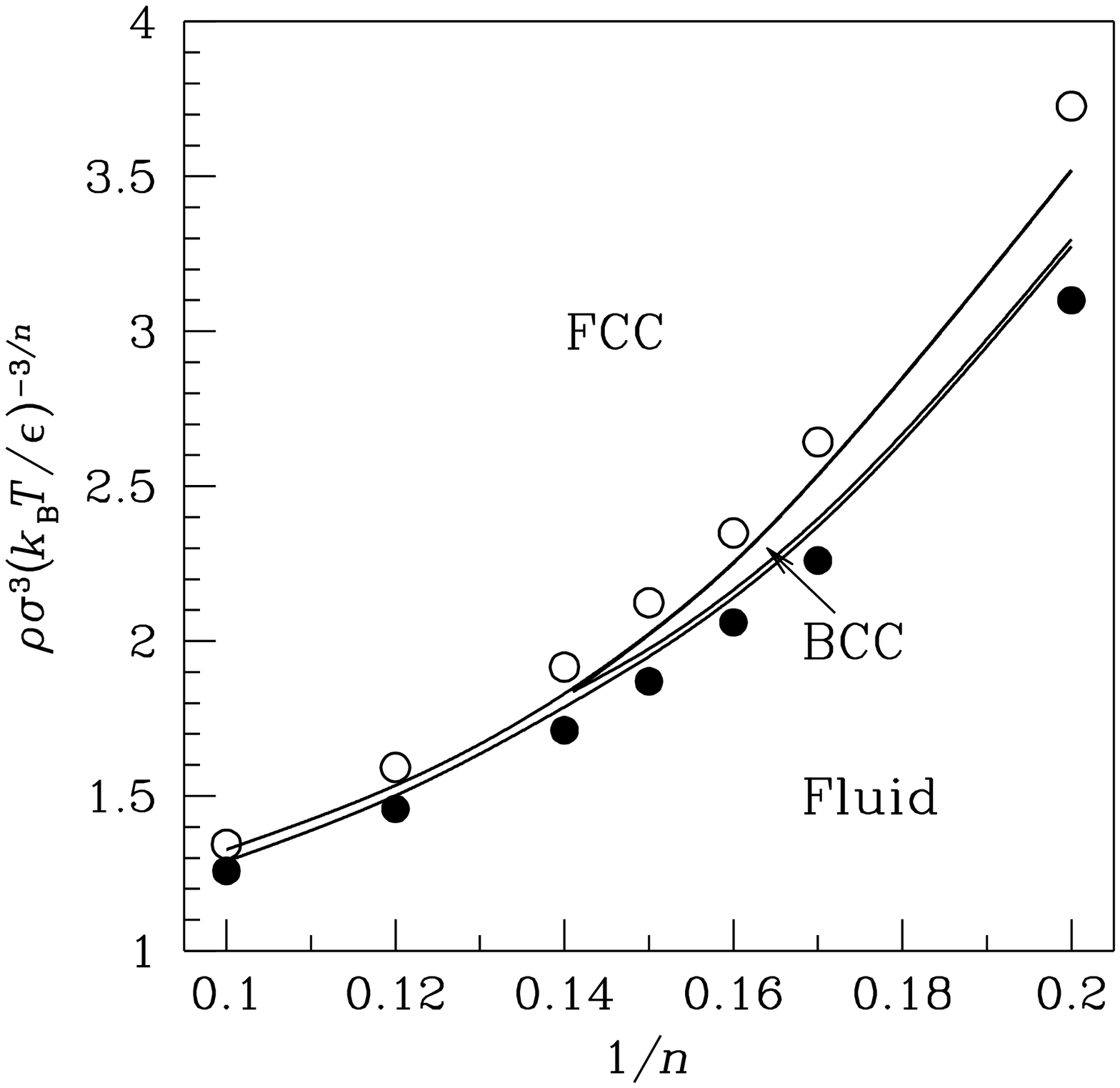}
\caption{} \label{fig2}
\end{figure}
%
%
\begin{figure}
\includegraphics[width=12cm,angle=0]{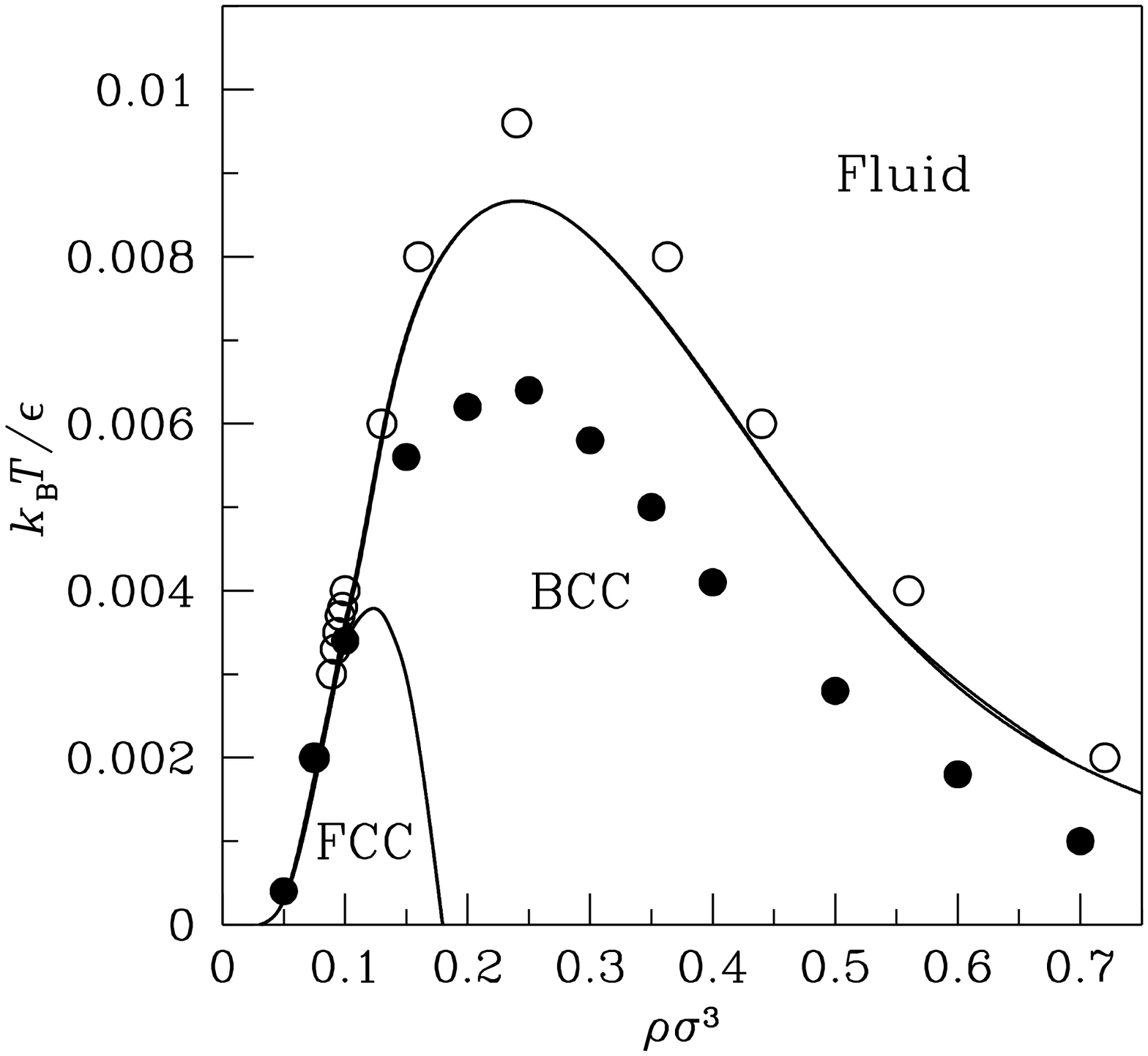}
\caption{} \label{fig3}
\end{figure}


\begin{thebibliography}{99}
\bibitem{Young} D.~A. Young, {\em Phase Diagrams of the Elements}, University of California Press, Berkeley (1991).
\bibitem{Monson} P.~A. Monson and D.~A. Kofke, Adv. Chem. Phys. {\bf 115}, 113 (2000).
\bibitem{Lowen} H. L\"owen, Phys. Rep. {\bf 237}, 249 (1994).
\bibitem{Prestipino1} S. Prestipino, F. Saija and P.~V. Giaquinta, Phys. Rev. E {\bf 71}, 050102(R) (2005).
\bibitem{Prestipino2} S. Prestipino, F. Saija and P.~V. Giaquinta, J. Chem. Phys. {\bf 123}, 144110 (2005).
\bibitem{Saija1} F. Saija and S. Prestipino, Phys. Rev. B {\bf 72}, 024113 (2005).
\bibitem{Lindemann} F.~A. Lindemann, Phys. Z. {\bf 11}, 609 (1910).
\bibitem{Hansen1} J.-P. Hansen and L. Verlet, Phys. Rev. {\bf 184}, 151 (1969).
\bibitem{Raveche} H. J. Ravech\'e, R. D. Mountain, W. B. Streett, J. Chem. Phys. {\bf 61}, 1970 (1974).
\bibitem{Giaquinta1} P.~V. Giaquinta and G. Giunta, Physica A {\bf 187}, 145 (1992).
\bibitem{Giaquinta2} P.~V. Giaquinta, {\em ``Entropy revisited: The interplay between ordering and correlations"}, in {\em
Highlights in the quantum theory of condensed matter} (Edizioni della Normale, Pisa, 2005; ISBN: 88-7642-170-X).
\bibitem{Meijer} E.~J. Meijer and D. Frenkel, J. Chem. Phys. {\bf 94}, 2269 (1991).
\bibitem{Hansen2} J.-P. Hansen and D. Schiff, Mol. Phys. {\bf 25}, 1281 (1973).
\bibitem{Nettleton} R.~E. Nettleton and M.~S. Green, J. Chem. Phys. {\bf 29}, 1365 (1958).
\bibitem{Saija2} F. Saija, S. Prestipino, and P.~V. Giaquinta, J. Chem. Phys. {\bf 113}, 2806 (2000).
\bibitem{Saija3} F. Saija, S. Prestipino, and P.~V. Giaquinta, J. Chem. Phys. {\bf 115}, 7586 (2001).
\bibitem{Saija4} F. Saija, G. Pastore, and P.~V. Giaquinta, J. Chem. Phys. {\bf 102}, 10368 (1998); F. Saija and P.~V. Giaquinta, J.
Phys. Chem. B {\bf 106}, 2035 (2002); F. Saija and P.~V. Giaquinta, J. Chem. Phys. {\bf 117}, 5780 (2002).
\bibitem{Costa} D. Costa, F. Micali, F. Saija, and P.~V. Giaquinta, J. Chem. Phys. {\bf 106}, 12297 (2002); D. Costa,
F. Saija, and P.~V. Giaquinta, J. Phys. Chem. B {\bf 107}, 9514 (2003).
\bibitem{Ross1} M. Ross and A.~K. McMahan, Phys. Rev. B {\bf 21}, 1658 (1980).
\bibitem{Ross2} M. Ross, R. Boehler, and P. S\"oderlind, Phys. Rev. Lett. {\bf 95}, 257801 (2005).
\bibitem{Vortler} H.~L. Vortler, I. Nezbeda, and M. Lisal, Mol. Phys. {\bf 92}, 813 (1997).
\bibitem{Lisal} M. Lisal, I. Nezbeda, and H. L. Vortler, Fluid Phase Equilibria {\bf 154}, 49 (1999).
\bibitem{Hoover}  W.~G. Hoover, M. Ross, K.~W. Johnson, D. Henderson, J.~A. Barker, and B.~C. Brown, J. Chem. Phys.
{\bf 52 }, 4931 (1970); W.~G. Hoover, S.~G. Gray, and K.~W. Johnson, J. Chem. Phys. {\bf 55}, 1128 (1971).
\bibitem{Laird} B.~B. Laird and A.~D.~J. Haymet, Mol. Phys. {\bf 75}, 71 (1992).
\bibitem{Agrawal} R. Agrawal and D.~A. Kofke, Phys. Rev. Lett. {\bf 74}, 122 (1995); Mol. Phys. {\bf 85}, 23 (1995).
\bibitem{Stillinger1} F.~H. Stillinger, J. Chem. Phys. {\bf 65}, 3968 (1976).
\bibitem{Lang}  A. Lang, C.~N. Likos, M. Watzlawek, and H. L\"owen, J. Phys.: Condens. Matter {\bf 12}, 5087 (2000).
\end{thebibliography}
\end{document}